\documentclass{article}

 \usepackage[sglblindworkshop, final]{neurips_2025}
\workshoptitle{Presented at the Amazon Trusted AI Symposium 2026, New York, NY, USA}

\usepackage[utf8]{inputenc} %
\usepackage[T1]{fontenc}    %
\usepackage{hyperref}       %
\usepackage{url}            %
\usepackage{booktabs}       %
\usepackage{amsfonts}       %
\usepackage{nicefrac}       %
\usepackage{microtype}      %
\usepackage{xcolor}         %
\usepackage{multirow}

\usepackage{svg}
\usepackage[most]{tcolorbox}
\usepackage{diagbox}
\usepackage{enumitem}
\usepackage{wrapfig}
\usepackage{caption}

\title{
Supercharging Federated Intelligence Retrieval
}

\author{%
  Dimitris Stripelis\thanks{Corresponding author: dimitris@flower.ai}, Patrick Foley, Mohammad Naseri, William Lindskog-Münzing,\\ \textbf{Chong Shen Ng, Daniel Janes Beutel, Nicholas D. Lane}\\
  Flower Labs
}

\begin{document}

\maketitle

\begin{abstract}
RAG typically assumes centralized access to documents, which breaks down when knowledge is distributed across private data silos. We propose a secure Federated RAG system built using Flower that performs local silo retrieval, while server-side aggregation and text generation run inside an attested, confidential compute environment, enabling confidential remote LLM inference even in the presence of honest-but-curious or compromised servers. We also propose a cascading inference approach that incorporates a non-confidential third-party model (e.g., Amazon Nova) as auxiliary context without weakening confidentiality.
\end{abstract}

\section{Introduction}

Retrieval-Augmented Generation (RAG) improves factual grounding and domain adaptability of LLMs by conditioning generation on external knowledge. However, RAG assumes centralized document access - a loose assumption that fails in practice, since organizational data is distributed across data silos and cannot be aggregated centrally due to regulatory, proprietary, or privacy constraints.

Federated RAG (FedRAG) mitigates this limitation by performing retrieval locally within each silo and aggregating the results on the server to construct a unified generation context~\cite{chakraborty2025federated}. This approach enables cross-silo knowledge access while preserving data locality, making it well-suited for sensitive domains such as healthcare and finance. However, most existing FedRAG pipelines offer limited security guarantees, as retrieved documents are exchanged in plaintext and can be exposed to honest-but-curious or compromised servers.

In this work, we present a secure FedRAG architecture built on Flower~\cite{beutel2020flower} that uses confidential computing to protect federated aggregation and generation within a trusted execution environment (TEE), with silo participation enforced via attestation and authorization~\cite{addison2024c}. 

Unlike prior FedRAG approaches based on fine-tuning~\cite{jung2025federated}, secure inference~\cite{addison2024c}, or encrypted retrieval and communication~\cite{addison2024c,zhao2024frag,mao2025privacy}, we enable confidential remote LLM inference via Flower Confidential Remote Compute (CRC)~\cite{flower2025intelligence}, a service running LLMs remotely with confidential computing. We also introduce a cascading inference approach that integrates a non-confidential third-party model (AWS Nova) as an auxiliary probabilistic knowledge source without compromising end-to-end confidentiality. In summary, our contributions are:

\begin{itemize}[nosep]
\item a secure Federated RAG system for cross-silo retrieval without data centralization
\item a confidential server-side aggregation and text generation using TEEs
\item a mixed-trust cascading inference mechanism integrating third-party models
\item an end-to-end protected remote LLM inference pipeline using Flower CRC
\end{itemize}

\section{Related Work \& Background}

\textbf{Federated Learning.}
Federated Learning (FL) enables collaborative model training across decentralized data silos without requiring raw data sharing~\cite{mcmahan2017communication}. While FL has been widely studied for model training and fine-tuning, its application to inference-time retrieval and generation has received comparatively less attention. Flower~\cite{beutel2020flower} provides a flexible framework for federated workloads beyond training, including secure aggregation and orchestration across heterogeneous clients.

\textbf{Confidential Compute.}
Confidential computing leverages hardware-based Trusted Execution Environments (TEEs) to protect code and data during execution, even from privileged system software~\cite{costan2016intel}. Recent systems integrate TEEs into distributed ML pipelines to protect training~\cite{aggarwal2020soteria}, inference~\cite{addison2024c}, and aggregation. However, many approaches focus on securing individual stages rather than end-to-end retrieval and generation pipelines. Flower CRC~\cite{flower2025intelligence} extends confidential computing to remote LLM inference, enabling large-scale models to be executed with strong confidentiality guarantees.

\textbf{Federated RAG.} 
Federated RAG (FedRAG) decentralizes retrieval by executing document search locally at each silo and aggregating retrieved results at a central server~\cite{chakraborty2025federated}. Prior work explores federated fine-tuning~\cite{jung2025federated}, encrypted retrieval and communication~\cite{zhao2024frag,mao2025privacy}, and secure aggregation~\cite{addison2024c}. However, most existing systems either expose retrieved documents in plaintext at the server or do not protect the generation step itself. In contrast, our approach combines federated retrieval with confidential aggregation and generation, while additionally supporting mixed-trust cascading inference with external LLM providers.

\textbf{Threat Model.}
We assume an honest-but-curious or compromised server operator and infrastructure provider, while trusting hardware-enforced TEEs and remote attestation mechanisms~\cite{costan2016intel, sabt2015trusted}. Clients (data silos) do not trust each other and never share raw documents, following standard federated learning assumptions~\cite{mcmahan2017communication, kairouz2021advances}. Network adversaries may observe, replay, or tamper with communication but cannot break standard cryptographic protections such as TLS~\cite{rescorla2018tls}. We follow prior confidential computing work in assuming the TEE implementation is correct and that side-channel attacks are out of scope~\cite{vanbulck2018foreshadow, chen2019sgxpectre}.

\section{Methods}
For every query, the server, operating inside a TEE, broadcasts the query to a set of verified clients/silos. Each client retrieves the top-$k$ relevant documents from its local private document store and returns them to the server along with their retrieval scores. The server aggregates the results, merges and re-ranks the retrieved documents, and constructs an augmented context that is passed to the LLM.

\begin{figure}[htbp]
    \centering
    \includegraphics[width=\textwidth]{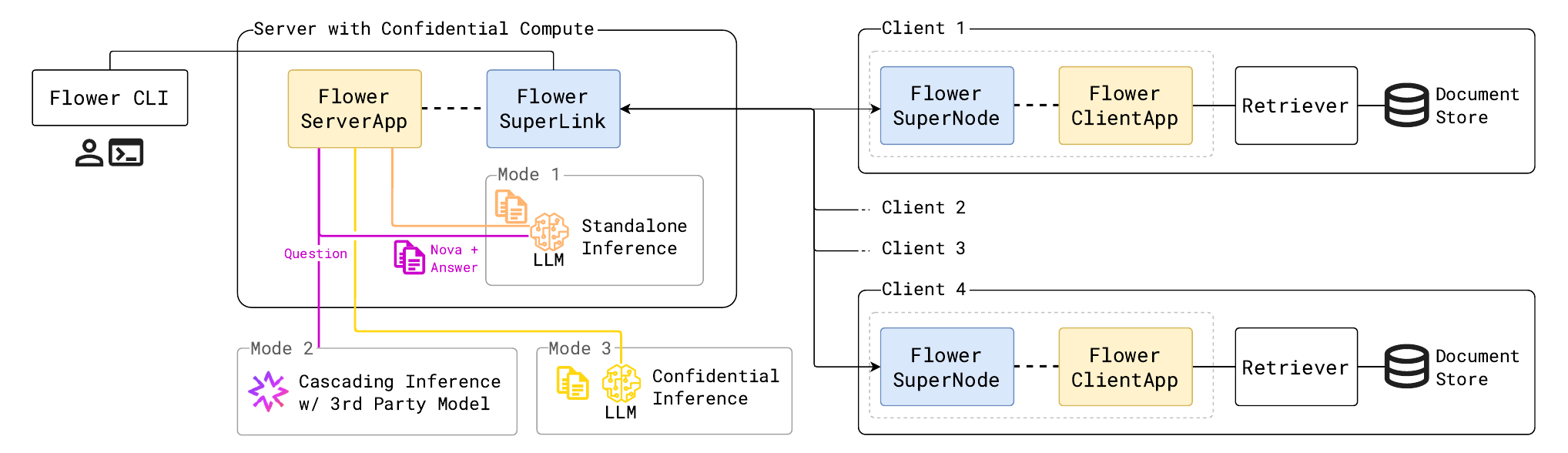}
    \caption{FedRAG with confidential server-side aggregation and various inference options.}
    \label{fig:fedrag_architecture}
\end{figure}

On the client side, retrieval is performed using a local FAISS index built over each client’s document corpus. The server merges the results into a single ranked list by sorting based on the retrieval score, computed as $1 / (k\-{RRF} + doc_{i} + 1)$, where $k\-{RRF}$ represents the value of the Reciprocal Rank Fusion and $doc_{i}$ the document index. Figure~\ref{fig:fedrag_architecture} illustrates the overall FedRAG pipeline.

In our setting, we evaluate three inference modes (server always operates inside a TEE):

\begin{itemize}[leftmargin=0.5cm, itemsep=0.01cm, topsep=0pt]
    \item \textbf{Standalone Inference:} The augmented query is processed by a server-hosted LLM, which generates a response using the retrieved documents as context.
    \item \textbf{Cascading Inference:} The server-hosted LLM generates the final response using both the retrieved documents and an auxiliary answer obtained from a non-confidential third-party provider, which is treated as an additional contextual source.    
    \item \textbf{Confidential Inference:} The augmented query is processed by a large-scale LLM running inside Flower CRC, ensuring that both prompts and context remain protected during inference.
\end{itemize}

\section{Evaluation}

\textbf{Experimental Settings.} We use the SmolLM 1.7B Instruct model as the server-hosted LLM. In the case of the cascading mode, the external 3rd party model was AWS Nova Micro, and in Flower CRC, we use the Qwen3 235B model. The entire server-side workload is CPU-bound; no GPU is employed to deploy SmolLM. In all experiments, the top-$k$ value is set to 8 and the reciprocal rank fusion parameter $k\-{RRF}$ to 60. The federation consists of four clients, each maintaining an independent document corpus: PubMed~\cite{nihPubMed} (23.9M snippets), StatPearls~\cite{nihStatPearls} (301k snippets), Textbooks~\cite{jin2021disease} (126k snippets), and Wikipedia~\cite{thakur2021beir} (29.9M snippets); sourced from the MedRAG Toolkit \cite{xiong2024benchmarking}. Evaluation is performed on the MIRAGE benchmark~\cite{xiong2024benchmarking}: PubMedQA (501 Y/N/maybe questions), BioASQ (619 Y/N questions), and MedQA (1272 A/B/C/D selection questions). Table~\ref{tab:evaluation} reports performance and mean end-to-end query time.

\begin{table}[htbp]
\centering
\footnotesize

\begin{tabular}{l|cc|cc|cc}
\toprule
\multirow{2}{*}{\textbf{\diagbox{Mode}{Dataset}}}
& \multicolumn{2}{c|}{\textbf{PubMedQA}}
& \multicolumn{2}{c|}{\textbf{BioASQ}}
& \multicolumn{2}{c}{\textbf{MedQA}}\\
\cmidrule(lr){2-3}\cmidrule(lr){4-5}\cmidrule(lr){6-7}
& \textbf{Time} & \textbf{Accuracy}
& \textbf{Time} & \textbf{Accuracy}
& \textbf{Time} & \textbf{Accuracy} \\
\midrule

Standalone Inference
& 43s & 0.32
& 44s & 0.81 
& 43s & 0.39 \\

Cascaded Inference
& 48s & 0.45
& 45s & 0.83
& 46s & 0.57 \\

Confidential Inference
& 25s & 0.47
& 26s & 0.82
& 25s & 0.78 \\

\bottomrule
\end{tabular}
\hspace{10cm} %
\caption{Mean query execution time and accuracy per benchmark dataset for each inference mode.}
\label{tab:evaluation}
\end{table}

\textbf{Results Analysis.} Compared to the standalone inference mode, cascading inference is especially compelling: even with the lightweight Amazon Nova Micro model, the additional context provided to SmolLM yields a $\sim$40\% and $\sim$46\% improvement on PubMedQA and MedQA, respectively, indicating that more powerful Nova Pro/Premier models are a promising direction for further evaluation. This result also highlights that despite the lower compute capabilities of CPU-based TEEs, their trust properties allow them to be used along with non-confidential external inference APIs to produce better performance.

Confidential inference with Qwen3 275B in Flower CRC achieves the highest accuracy and lowest latency, consistent with running a large capacity model on an NVIDIA H100 GPU. In contrast, CPU-bound generation in SmolLM dominates latency in standalone and cascading modes. Finally, confidential inference benefits from receiving encrypted aggregated documents directly from the Server TEE, with documents only decrypted within Flower CRC's hardened TEE. Overall, both the cascading and confidential inference modes outperform standalone while securely combining context.

\section{Conclusion}
We present a secure Federated RAG architecture built on Flower that preserves data locality and protects aggregation and generation via confidential computing. Experiments show that cascading inference using a lightweight non-confidential third-party model for probabilistic contextual augmentation improves performance, highlighting the practicality of CPU-based TEEs for mixed-trust deployments. Leveraging high-capacity GPU-backed models inside hardened TEEs, confidential inference with Flower CRC achieves higher accuracy and lower latency, enabling a flexible, privacy-preserving RAG design that supports both confidential and third-party inference endpoints.

\newpage
\bibliographystyle{unsrt}
\bibliography{bibliography.bib}

\newpage

\appendix

\section{Corpus Specifications}
The current implementation of document retrieval for the FAISS index is built with IndexFlatL2 and uses the $faiss.METRIC_{L2}$ metric for measuring dissimilarity distance; the lower the score for a retrieved document is, the closer semantically the document is to the received query. Table~\ref{tbl:data-corpus-specs} gives an overview of the specifications of each corpus.

\begin{table}[htbp]
\centering
\begin{tabular}{lrrrr}
\toprule
\textbf{Corpus} & \textbf{Size} & \textbf{\#Snippets} \\
\midrule
PubMed      & 70GB & 23.9M \\
StatPearls & 2GB  & 301k\\
Textbooks   & 209MB & 126k\\
Wikipedia   & 44GB & 29.9M\\
\bottomrule
\end{tabular}
\hspace{10cm} %
\caption{Data Corpus Specifications.}
\label{tbl:data-corpus-specs}
\end{table}

\section{Question \& Prompt Sample}
When all documents are retrieved from the distributed data silos, they are used as context to answer the user query. A typical example is shown in the color box below, which presents the model's role and instruction, the relevant documents/snippets retrieved from the document store of the participating silos, the original question, and the answering options.

\begin{tcolorbox}[
  title={Sample Question Prompt},
  colback=gray!10,
  colframe=gray,
  sharp corners
]

As an expert doctor in clinical science and medical knowledge, can you tell me if the following statement is correct? Answer yes, no, or maybe.
\\\\
\textbf{Here are the relevant documents:}

Document 1: Another set of strategies for improving quality involves changing the systems of care...

Document 2: To enhance interprofessional healthcare team outcomes...

Document 3: and surgical morbidity...

Document 4: Social workers focus on educating families...

Document 5: a supervisory capacity while creating a culture of safety...

Document 6: a hospital setting. Many doctors prefer working in an ASC...

Document 7: leaders using an empowering style may improve process...

Document 8: Coordinated care involves seamless transitions...
\\\\
\textbf{Question:}
Does a dedicated discharge coordinator improve the quality of hospital discharge?
\\\\
\textbf{Options:}
A) yes  
B) no  
C) maybe
\\\\
Answer only with the correct option
\label{box:sample-question-prompt}
\end{tcolorbox}

\end{document}